\documentclass[prx,aps,floats,twocolumn]{revtex4}
\usepackage[dvips]{graphicx}
\usepackage{amssymb}
\usepackage{amsmath}
\usepackage{physics}
\usepackage{natbib}

\usepackage{color}

\begin{document}

\title{Conformal Quasicrystals and Holography}

\author{Latham Boyle$^1$, Madeline Dickens$^{2}$ and Felix Flicker$^{2,3}$} 

\affiliation{$^{1}$Perimeter Institute for Theoretical Physics, Waterloo, Ontario N2L 2Y5, Canada, N2L 2Y5 \\
$^{2}$Department of Physics, University of California, Berkeley, California 94720, USA \\
$^{3}$Rudolf Peierls Centre for Theoretical Physics, University of Oxford, Department of Physics, Clarendon Laboratory, Parks Road, Oxford, OX1 3PU, United Kingdom}

\begin{abstract}
Recent studies of holographic tensor network models defined on regular tessellations of hyperbolic space have not yet addressed the underlying discrete geometry of the boundary. We show that the boundary degrees of freedom naturally live on a novel structure, a \textit{conformal quasicrystal}, that provides a discrete model of conformal geometry. We introduce and construct a class of one-dimensional conformal quasicrystals, and discuss a higher-dimensional example (related to the Penrose tiling).  Our construction permits discretizations of conformal field theories that preserve an infinite discrete subgroup of the global conformal group at the cost of lattice periodicity.
\end{abstract}
\maketitle 

\section{Introduction}

A central topic in theoretical physics over the past two decades has been holography: the idea that a quantum theory in a bulk space may be precisely dual to another living on the boundary of that space. The most concrete and widely-studied realization of this idea has been the AdS/CFT correspondence \cite{Maldacena:1997re, Witten:1998qj, KaplanNotes}, in which a gravitational theory living in a ($d+1$)-dimensional negatively-curved bulk spacetime is dual to a non-gravitational theory living on its $d$-dimensional boundary. Over the past decade, investigations of this duality have yielded mounting evidence that spacetime and its geometry can in some sense be regarded as emergent phenomena, reflecting the entanglement pattern of some other underlying quantum degrees of freedom \cite{VanRaamsdonk:2010pw}.  

Recently, physicists have been interested in the possibility that {\it discrete} models of holography would permit them to understand this idea in greater detail, by bringing in the tools of condensed matter physics and quantum information theory --- much as lattice gauge theory led to conceptual and practical progress in understanding gauge theories in the continuum~\cite{Wilson:1974sk}. Recent years have seen a surge of interest in discrete models of holography based on tensor networks (TNs). Such holographic TNs relate the quantum degrees of freedom (DOF) living on a discretized $(d+1)$-dimensional hyperbolic space to corresponding DOF on the discretized $d$-dimensional boundary of that space. They bridge condensed matter physics, quantum gravity, and quantum information. 

In condensed matter physics, holographic TNs provide a computationally efficient description of the highly-entangled ground-states of quantum many-body systems (particularly scale invariant systems or systems near their critical points). The entanglement pattern is described by tensors living on an emergent discrete hyperbolic geometry in one dimension higher \cite{Vidal:2007hda, Vidal:2008zz, Vidal:Intro, Pfeifer:2008jt, EvenblyVidalTNgeometry, Orus:2014poa}. On the quantum gravity side, 
the TNs are a kind of UV regulator of the physics in the bulk, and 
a proposed way to represent the fact that spacetime and its geometry may be regarded as emergent phenomena, reflecting the entanglement pattern of some other underlying quantum degrees of freedom \cite{Swingle:2009bg, Swingle:2012wq, Qi:2013caa, MolinaVilaplana:2012nz, Pastawski:2015qua, Hayden:2016cfa, Czech:2015kbp, Han:2016xmb, Qi:2017ohu, Evenbly:2017hyg, Osborne:2017woa, Bao:2017qmt, Jahn:2017tls}. Meanwhile, quantum information theory provides a unifying language for these studies in terms of entanglement, quantum circuits, and quantum error correction \cite{NielsenChuang}.  

These investigations have gradually clarified our understanding of the discrete geometry in the bulk. There has been a common expectation, based on an analogy with AdS/CFT \cite{Maldacena:1997re, Witten:1998qj, KaplanNotes}, that TNs living on discretizations of a hyperbolic space define a lattice state of a critical system on the boundary and vice-versa. Initially, this led Swingle to interpret Vidal's Multiscale Entanglement Renormalization Ansatz (MERA) as a discretized version of a time-slice of the AdS geometry \cite{Vidal:2008zz, Swingle:2009bg}. However, MERA has a preferred causal direction, while any discretization of a spacelike manifold should not.  To fix this issue, it was proposed that MERA was related, first to a different object known as the kinematic space of AdS \cite{Czech:2015kbp, Czech:2015qta} and, more recently, to a light-like geometry \cite{Milsted:2018san}. 

Meanwhile, new TN models ({\it e.g.~}the holographic quantum error-correcting codes of \cite{Pastawski:2015qua}, the hyper-invariant networks of \cite{Evenbly:2017hyg}, or the matchgate networks of \cite{Jahn:2017tls}) were introduced to more adequately capture AdS. The key feature of these new TNs is that they live on regular tilings of hyperbolic space~\cite{CoxeterHyperbolic}. The symmetries of such a tessellation form an infinite discrete subgroup of the continuous symmetry group of the AdS time-slice, much as the symmetries of an ordinary 3D lattice or crystal form an infinite discrete subgroup of the continuous symmetry group of 3D Euclidean space.  Physically, such TNs represent a discretization of the continuous bulk space, much as a crystal represents a discretization of a continuum material. In this paper, we restrict our attention to TNs of this type.

Despite progress in describing the discrete {\it bulk} geometry, the question of which discrete geometric spaces are suitable for the {\it boundary} has received little attention. In a discrete model of the AdS/CFT correspondence, we expect to be able to construct the discrete boundary geometry entirely from the data of the bulk tessellation, in analogy with the continuum case, where it is well-known that the data of any asymptotically AdS spacetime define a conformal manifold on its boundary \cite{Witten:1998qj}. While this expectation seems natural, we are unaware of a corresponding discrete construction in the literature. In condensed matter physics, the discrete boundary geometry is typically assumed to be a periodic lattice; but, as we observe here, there is no natural way in which an ordinary regular tessellation of hyperbolic space defines a periodic lattice on its boundary, and no natural way for the discrete symmetries of the bulk to act on such a boundary lattice \footnote{Note that, in this paper, as in \cite{Pastawski:2015qua, Evenbly:2017hyg, Jahn:2017tls}, we focus on {\it ordinary} regular 
$\{p,q\}$ tesselations ({\it i.e.}\ tesselations where each tile is a regular $p$-gon, $q$ such tiles meet at each vertex, and $p$ and $q$ are both finite), 
since these represent the natural discretizations of theories that are local in the bulk.  Another fascinating possibility is to relax this locality restriction and consider 
$\{p,\infty\}$ 
tilings of the hyperbolic plane in which infinitely many tiles meet at each vertex; in this case, tile edges are infinitely long, and the tile vertices live at infinity, on the boundary of the hyperbolic plane, where they {\it can} then induce a periodic structure at infinity.  Ref.~\cite{Osborne:2017woa} contains a detailed exploration of this latter possibility and, in particular, in this context presents an interesting toy model in which the discrete symmetries in the bulk may be related to a discretization of the conformal group of the boundary.  
The dual cases $\{\infty,q\}$ consist of regular polygons with an infinite number of sides (apeirogons), with $q$ meeting at each vertex. When $q$ is a prime $p$, the boundary theory is then naturally described by the $p$-adic numbers~\cite{Brekke:1993gf}.}
\footnote{Yet another possibility is to replace the regular tiling in the bulk by a {\it random} tensor network as in \cite{Hayden:2016cfa}.  In AdS/CFT, a key role is played by the isometry group of the bulk AdS space, and a regular tiling has the advantage that it preserves a large discrete subgroup of this symmetry.  However, for some purposes, a random tensor network may also have certain advantages, and it would be interesting to investigate the mathematical structure that it induces on the boundary.  If the random tensor network in the bulk respects the properties of the regular tiling in a statistical sense, then the random structure induced on the boundary should also respect the properties of the corresponding conformal quasicrsytal in a statistical sense, but we leave the investigation of this topic to future work.}.  
Thus we expect that, in implementing a discrete version of 
AdS/CFT,
it will be natural to replace the periodic boundary lattice with a different discrete object.  Such a replacement is reasonable, from a Wilsonian viewpoint, as the choice of underlying lattice becomes irrelevant at a critical point~\cite{luck1993classification}.

\begin{figure}[t]
\includegraphics[width=.48\textwidth]{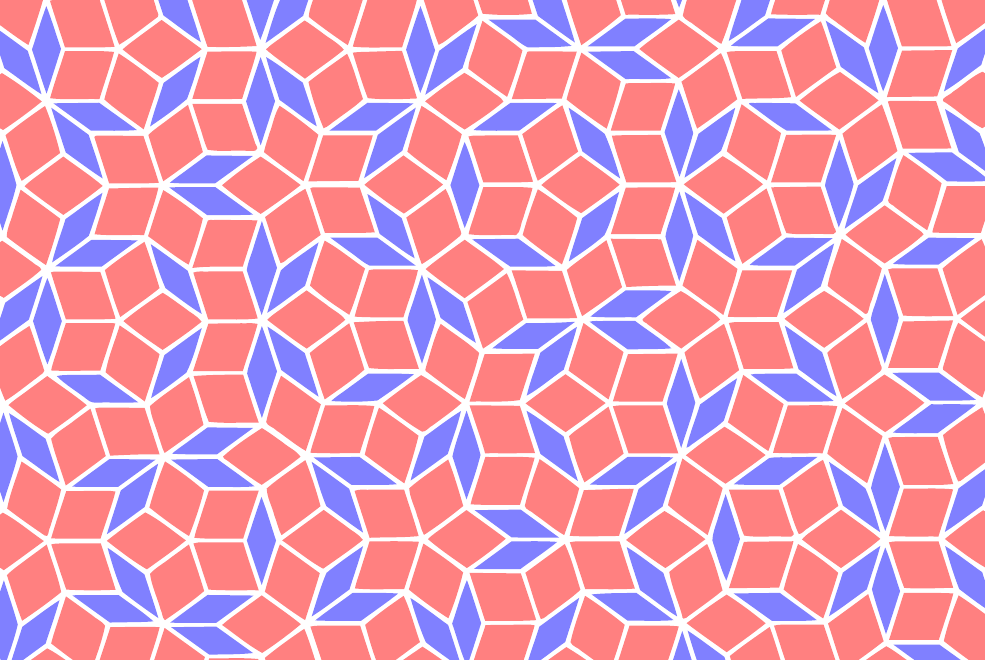}
\caption{The Penrose tiling \cite{BaakeGrimm}. The two different tiles (red and blue) cover the entire Euclidean plane in an aperiodic manner without gaps. Centers of approximate five-fold symmetry can be seen. The tiling can be created through a cut-and-project method from a five-dimensional hypercubic lattice or a four-dimensional $A_4$ lattice, or by applying `inflation rules' to a finite-size starting configuration~\cite{Senechal,BaakeGrimm,Janot}.
}
\label{Penrose}
\end{figure}

In this paper, we argue that the DOF on the boundary of a regular tessellation of hyperbolic space naturally live on a remarkable structure -- a \textit{conformal quasicrystal} (CQC) -- built entirely from the data of the bulk tessellation. 
This is a new clue about the type of boundary theory that should appear in a discrete version of holography.   
Far from being a simple periodic lattice, each CQC {\it locally} resembles a self-similar quasicrystal \cite{Senechal, Janot, BaakeGrimm, Boyle:2016sjm, Boyle:2016iey} (like the Penrose tiling~\cite{Gardner}, shown in Fig.~\ref{Penrose}) and possesses discrete symmetries worthy of a discretization of a conformal manifold.   In this paper, we focus on the $d=1$ case, but with a view towards higher dimensions (which we discuss briefly near the end).  We propose to use our framework to construct discretizations of CFTs that preserve an infinite discrete subgroup of the global conformal group at the cost of exact discrete translation invariance (which is replaced by quasi-translational invariance). We end with suggestions for future research, and a discussion of how our results will hopefully provide an important clue in the ongoing effort to formulate a discrete version of holography, and also lead to an improved analytical and numerical understanding of the structure of condensed matter systems at their critical points, and other conformally invariant systems.

\section{Quasicrystals in $d=1$.} 

We begin by briefly introducing self-similar quasicrystals (SSQCs) in $d = 1$ (see Refs.~\cite{Senechal, Janot, BaakeGrimm, Boyle:2016sjm, Boyle:2016iey} for further details). For general $d$, an SSQC may be thought of as a special quasiperiodically-ordered set of discrete points in Euclidean space (like the vertices in an infinite 2D Penrose tiling, shown in Fig.~\ref{Penrose}). The same SSQC can be constructed in two very different ways: either (i) by a Ôcut-and-projectÕ method based on taking an irrationally-sloped slice through a periodic lattice in a higher-dimensional Euclidean space, and projecting nearby points onto the slice; or (ii) by recursive iteration of an appropriate substitution (or `inflation') rule. Restricting to $d=1$ in the following sections, we exploit the second (inflation) perspective to develop a novel construction whereby 1D SSQCs emerge on the boundary of a lattice in 2D \textit{hyperbolic} space.

Let $\Pi$ be a (possibly infinite) string of two letters, $\alpha$ and $\beta$; and let $s_{\alpha}(\alpha,\beta)$ 
and $s_{\beta}(\alpha,\beta)$ be two finite strings of $\alpha$s and $\beta$s.  We can act on $\Pi$ with the {\it inflation rule}
\begin{align}
	\tau:\;(\alpha,\beta)\mapsto (s_\alpha(\alpha,\beta),s_\beta(\alpha,\beta))
\end{align} 
which replaces each $\alpha$ or $\beta$ in $\Pi$ by the string $s_{\alpha}(\alpha,\beta)$ or $s_{\beta}(\alpha,\beta)$, respectively, to obtain a new string $\Pi'$.  The inverse map is the corresponding {\it deflation rule}:
\begin{align}
	\tau^{-1}:\;(s_\alpha(\alpha,\beta),s_\beta(\alpha,\beta))\mapsto (\alpha,\beta).
\end{align} 
Note that, while the inflation rule has a well-defined action on any string $\Pi$ of $\alpha$s and $\beta$s, the corresponding deflation map only has a well-defined action on a string $\Pi$ that may be uniquely partitioned into substrings of the form $s_{\alpha}(\alpha,\beta)$ and $s_{\beta}(\alpha,\beta)$.  

Any inflation rule $\tau$ induces a matrix $M_\tau$ that encodes the growth of $N_{\alpha}$ and $N_{\beta}$ (the number of $\alpha$s and $\beta$s, respectively, in the string). For example, under $\tau:(\alpha,\beta)\mapsto (\alpha\beta\alpha,\alpha\beta\alpha\beta\alpha)$, which corresponds to inflation rule $3b$ in Table 1 of \cite{Boyle:2016sjm}, we have
\begin{align}
  \label{tau_QC_example}
  \mqty(N_{\alpha}' \\ N_{\beta}')=M_\tau\mqty(N_{\alpha} \\ N_{\beta}),\;\;\;M_\tau=\mqty(2&3\\1&2).
\end{align}
If $\lambda$ is the largest eigenvalue of $M_{\tau}$ ($\lambda=3+\sqrt{2}$), we can represent $\Pi$ geometrically as a sequence of 1D line segments or `tiles' of length $L_{\alpha}$ and $L_{\beta}$, where $(L_{\alpha},L_{\beta})$ is the corresponding {\it left} eigenvector of $M_{\tau}$. If $N_{k}^{\alpha}$ and $N_{k}^{\beta}$ denote the number of $\alpha$s and $\beta$s after $k$ successive applications of $\tau$ to some finite initial string, then in the limit $k\to\infty$, $(N_{k}^{\alpha},N_{k}^{\beta})$ is the corresponding {\it right} eigenvector of $M_{\tau}$ and grows like $(N_{k+1}^{\alpha},N_{k+1}^{\beta})=\lambda(N_{k}^{\alpha},N_{k}^{\beta})$.

Two strings $\Pi$ and $\Pi'$ are \textit{locally isomorphic} (or {\it locally indistinguishable}) if every finite substring contained in one is also contained in the other so it would be impossible to distinguish them by inspecting any finite segment.

We say that $\Pi$ is a {\it $\tau$-quasicrystal} if it is equipped with an inflation rule $\tau$ acting on the two interval types $\alpha$ and
$\beta$ \footnote{Strictly speaking, this is a $\tau$-quasicrystal \textit{of degree two}, since $\tau$ acts on two interval types; if $\tau$ acts on $n$ interval types we would have a $\tau$-quasicrystal of degree $n$.} such that: (i) the matrix $M_{\tau}$ has determinant $\pm1$, with its largest eigenvalue $\lambda$ given by an irrational Pisot-Vijayaraghavan (PV) number; and (ii) the $k$-fold deflation map $\tau^{-k}$ is well-defined on $\Pi$ for all positive integers $k$.  Condition (i) ensures that $\Pi$ is {\it quasiperiodic}, {\it crystalline} (in the sense that the Fourier representation of its density profile exhibits delta function diffraction peaks like a crystal), and has $\tau^{-1}$ as the unique deflation rule that inverts $\tau$ \cite{Senechal, BombieriTaylor}.  Condition (ii) implies that a particular $\tau$-quasicrystal is locally isomorphic to every other $\tau$-quasicrystal and, in particular, is locally isomorphic to its own descendants under inflation.  In this sense, every $\tau$-quasicrystal is {\it self-similar} under inflation by $\tau$.  If a $\tau$-quasicrystal $\Pi$ is mapped to {\it itself} by ($k$ successive applications of) $\tau$, we say it is ($k$-fold) {\it self-same}, an even stronger scale-invariance property than self-similarity.

\section{Unit cell assignment on the hyperbolic disk}
\begin{figure}[t]
\includegraphics[width=.46\textwidth]{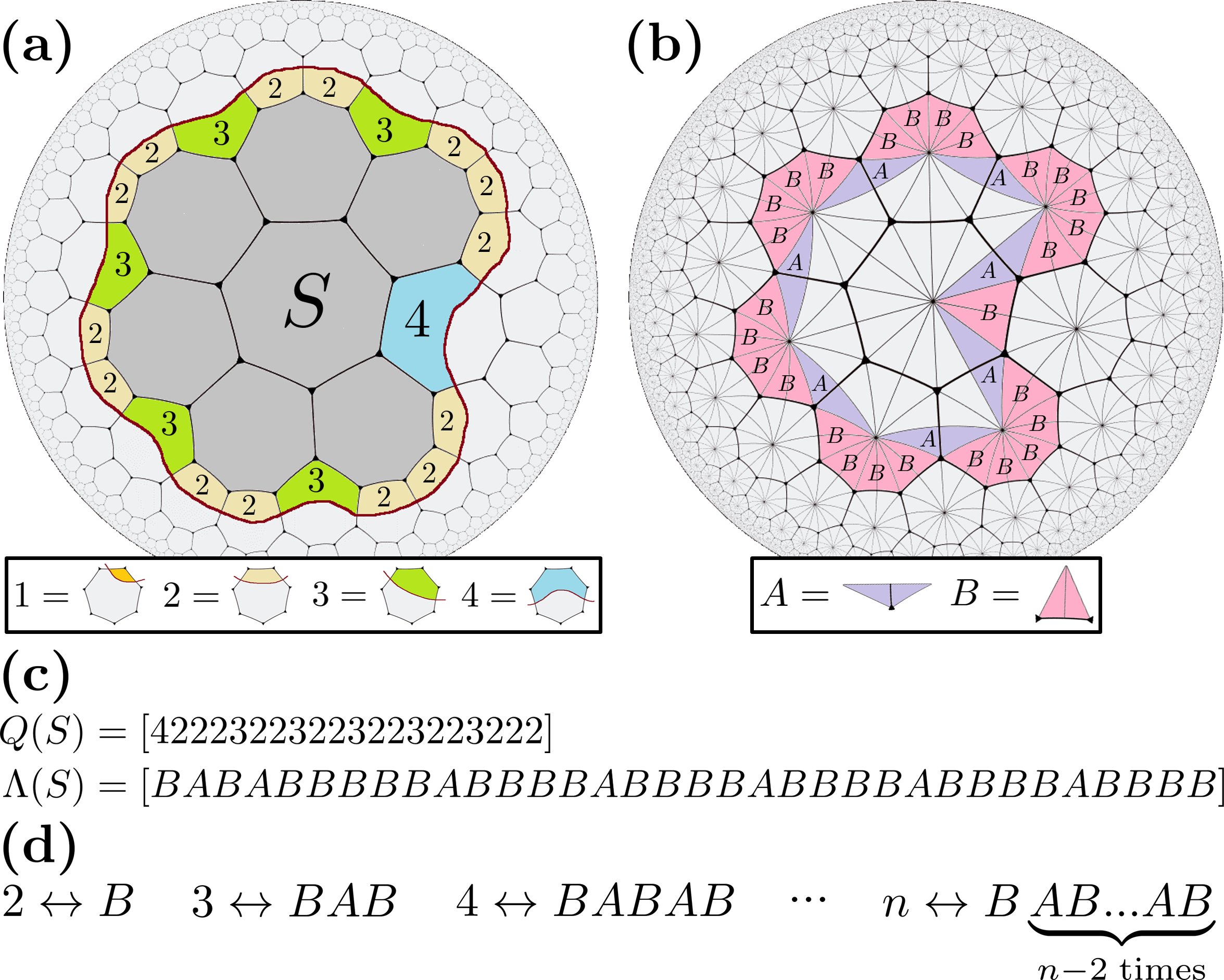}
\caption{Illustration of unit cell and letter assignment on the boundary of a simply-connected tile set $S$, in the $\qty{7,3}$ tiling. {\bf(a)} Unit cell assignment. {\bf(b)} Letter assignment. {\bf(c)} Cyclically ordered strings of unit cells and letters. {\bf(d)} Correspondence between unit cells and letters in the $\qty{7,3}$ case.}
\label{Assignment}
\end{figure} 

A regular $\{p,q\}$ tessellation is constructed from regular $p$-sided polygons (with $p$ geodesic edges), where $q$ such polygons meet at each vertex.  Such a tessellation exists for any $p,q>2$: when $(1/p)+(1/q)$ is greater than, equal to, or less than $1/2$, respectively, it is a tessellation of the sphere, Euclidean plane, or hyperbolic plane \cite{CoxeterHyperbolic, CoxeterRegularPolytopes}. 

Consider a regular $\qty{p,q}$ tessellation of the hyperbolic plane, $\frac{1}{p}+\frac{1}{q}<\frac{1}{2}$. We will focus in this paper on the case where $p$ and $q$ are both finite. Let $S$ (the `seed') be a finite, simply-connected patch of tiles; and let it also be {\it convex}, which here means that, if $\theta_c$ is the interior angle of $S$ at a vertex $c$, then: $\theta_{c}<2\pi(1-1/q)$ for each boundary vertex (when $q>3$); or $\theta_a+\theta_b<4\pi(1-1/q)$ for all nearest-neighbor pairs $\ev{a,b}$ of boundary vertices (when $q=3$).  On the boundary of $S$, we can assign different unit cell types between which the DOF live. The assignment procedure, illustrated in Fig.~\ref{Assignment}a, is the following. Consider the union $U(S)$ of all tiles that share a vertex with the boundary of $S$ (excluding the tiles in $S$ itself).  In the interior of $U(S)$, we can draw a curve that crosses every tile of $U(S)$ exactly once.  This curve partitions every tile into two pieces, one contained in the interior of the curve and one in the exterior.  If the interior part of the tile has $n$ vertices, then we call it a type-$n$ unit cell.  Once every cell on the boundary of $S$ has been labeled in this manner, we can write down  a corresponding string $Q(S)$ of cell types as in Fig.~\ref{Assignment}c.  These strings are inside square brackets to emphasize that they are cyclically ordered.  When $S$ is convex, only two cell types appear on its boundary (type-1 and type-2 when $q>3$, and type-2 and type-3 when $q=3$); but since the seed $S$ in Fig.~\ref{Assignment}a is {\it not} convex, a type-4 cell also appears.

In a concrete holographic TN model, a tensor is placed on each vertex of $S$ and contractions are performed between nearest-neighbor tensors along the tessellation edges. Extra tensors can be added to the edges, as in the hyperinvariant networks of Evenbly \cite{Evenbly:2017hyg}, or bulk ancilla DOF to the vertex tensors, as in the holographic quantum error correcting codes of Ref.~\cite{Pastawski:2015qua}. The boundary DOF are defined on legs placed on the edges of the tessellation emanating from $S$ which are cut by the partitioning curve in Fig.~\ref{Assignment}a. Thus, the distance in the network between two adjacent DOF on the boundary of $S$ depends on the type of unit cell between them. In this sense, the cell types defined here behave analogously to the interval types $L_\alpha$ and $L_\beta$ between DOF placed on the vertices of a $d=1$ quasicrystal. 

There is an alternative way to label the boundary of $S$ using only two symbols, that treats the $q=3$ and $q>3$ cases more uniformly, and continues to apply even when $S$ is nonconvex. Decompose every tile into the fundamental domains of the tiling symmetry group, which are right triangles of angles $\pi/2$, $\pi/q$, and $\pi/p$.  This decomposition is shown in Fig.~\ref{Assignment}b.  Now consider the set of such right triangles in the interior of $S$ that share a vertex with the boundary of $S$. Within this set we identify and label two types of isosceles triangles -- $A$ and $B$. Each $A$ shares one vertex and no edges with the boundary of $S$, and each $B$ shares two vertices and one edge (see Fig.~\ref{Assignment}b). This procedure, \textit{letter assignment}, produces a cyclically-ordered string $\Lambda(S)$, distinct from $Q(S)$, containing only $A$s and $B$s (see Fig.~\ref{Assignment}c).  

There is a way to extract $Q(S)$ from $\Lambda(S)$ and vice-versa by mapping each type-$n$ unit cell to a string of $A$s and $B$s. This correspondence is
\begin{align*}
	1&\leftrightarrow A^{-1}&\\
	2&\leftrightarrow A^{\frac{q-3}{2}}BA^{\frac{q-3}{2}}&\\
	3&\leftrightarrow A^{\frac{q-3}{2}}BA^{q-2}BA^{\frac{q-3}{2}}&\\
	&\,\,\vdots&\\
	n&\leftrightarrow A^{\frac{q-3}{2}}(BA^{q-2})^{n-2}BA^{\frac{q-3}{2}},&
\end{align*}
where $n\le p$. Here, if $f(A,B)$ is any finite substring then $(f(A,B))^x$ denotes its $x$-fold repetition whenever $x\in\mathbb{Z}_{\ge 0}$, with $(f(A,B))^0$ the empty substring. Fractional and negative exponents behave in the usual way, \textit{e.g.} $A^{x}A^{-y}=A^{x-y}$ for all $x,y\in\mathbb{Q}$.

\section{Inflation rules on the hyperbolic disk} 
\begin{figure}[t]
\includegraphics[width=.48\textwidth]{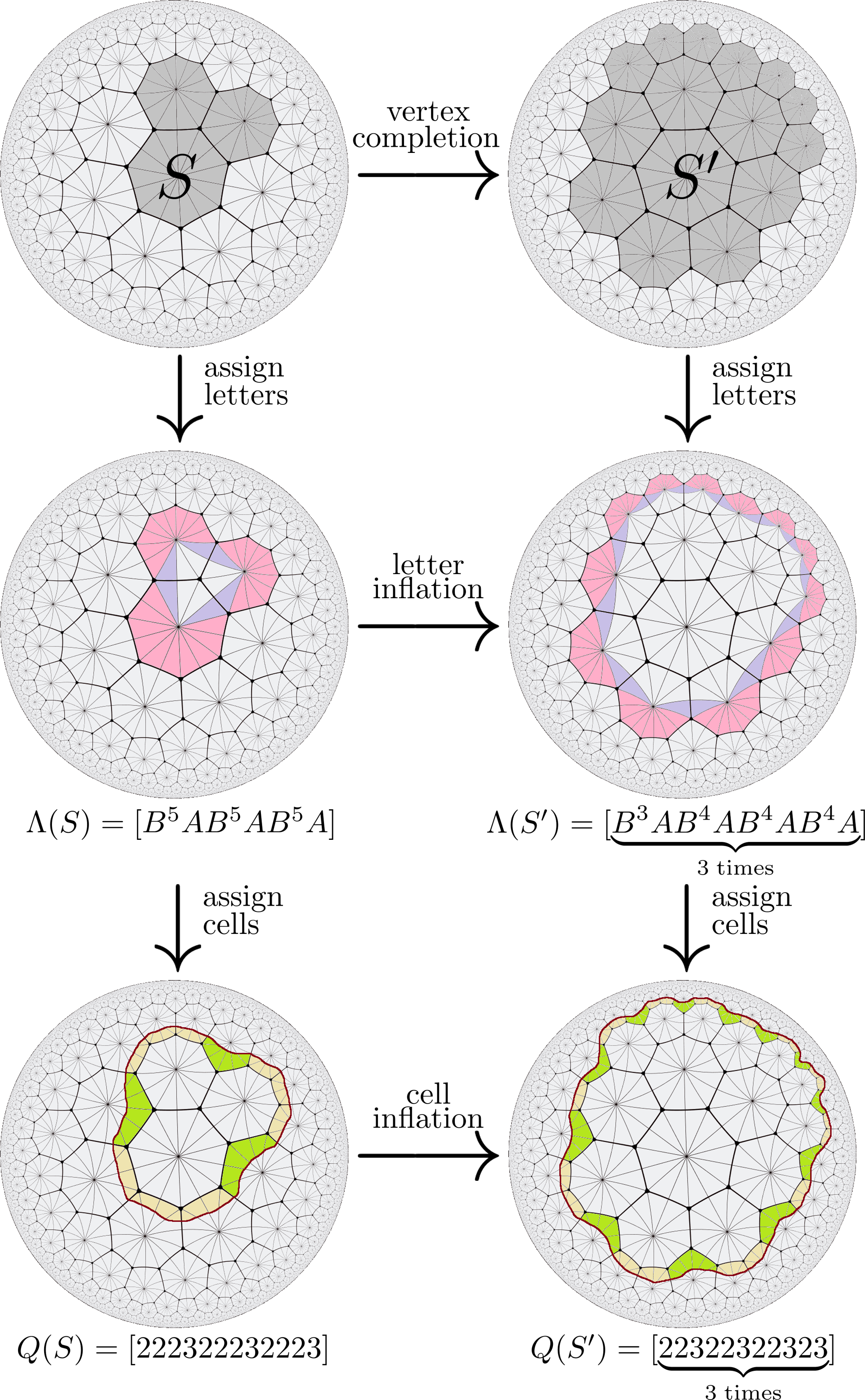}
\caption{Demonstration of how vertex completion induces inflation rules on boundary letters and unit cells, for the $\qty{7,3}$ tiling. The induced letter inflation rule $(A,B)\mapsto (A^{-1}B^{-5},B^4A)$ is the same for \textit{every} tile set $S$, convex or not. When $S$ is convex, only type-2 and type-3 unit cells appear and there is an induced unit cell inflation rule $(2,3)\mapsto(223,23)$. Note that vertex completion preserves convexity and that every tile set $S$ becomes convex after finitely many vertex completions. }
\label{inflation}
\end{figure} 

We can append the ring of tiles $U(S)$ to the set $S$ to obtain a new larger set $S'=S+U(S)$, as in Fig.~\ref{inflation}.  Accordingly, we add a layer of tensors to our TN to obtain a holographic TN model on $S'$ whose DOF are defined between the unit cells on the boundary of $S'$.
The vertices that were formerly on the boundary ({\it i.e.}\ the ones that were not completely surrounded by the tiles of $S$) are now in the interior ({\it i.e.}\ are completely surrounded by the tiles of $S'$). We dub this procedure \textit{vertex completion}. 

Vertex completion induces a mapping $\tau_\Lambda(p,q):\Lambda(S)\rightarrow \Lambda(S')$. Remarkably, $\tau_\Lambda(p,q)$ is an inflation rule on $A$ and $B$ and is the same for every $S$. If we define the stringlets
\begin{equation} 
  s_{L}:=A^{1/2}B^{(p-2)/2},\qquad s_{R}:=B^{(p-2)/2}A^{1/2}, 
\end{equation}
then $\tau_\Lambda(p,q)$ can be written 
\begin{subequations}
  \label{AB_symmetric}
  \begin{eqnarray}
    A&\mapsto&(s_{L}s_{R})^{-1}=A^{-1/2}B^{2-p}A^{-1/2}, \\
    B&\mapsto&\left\{\begin{array}{cl}
    s_{L}(s_{R}s_{L})^{\frac{q-3}{2}}B^{-1}(s_{R}s_{L})^{\frac{q-3}{2}}s_{R} & \;\;(q\;{\rm odd}),\quad \\
    (s_{R}s_{L})^{\frac{q-2}{2}}B^{-1}(s_{R}s_{L})^{\frac{q-2}{2}} & \;\;(q\;{\rm even}).\quad \end{array}\right.\quad
  \end{eqnarray}
\end{subequations}
This version of the substitution rule is the canonical one, in the sense that it respects the symmetry of the parent tile ($A$ or $B$) under reflection about its midline, but another equivalent version (which is asymmetric) is simpler and more convenient:
\begin{align}
  \label{AB_asymmetric}
		A\mapsto A^{-1}B^{2-p},\;\;\;
		B\mapsto B^{p-3}A(B^{p-2}A)^{q-3}.
\end{align}
Note that two substitution rules $\{\alpha,\beta\}\mapsto\{s_{\alpha},s_{\beta}\}$ and $\{\alpha,\beta\}\mapsto\{\bar{s}_{\alpha},\bar{s}_{\beta}\}$
are equivalent if $\bar{s}_{\alpha}=u s_{\alpha}u^{-1}$ and $\bar{s}_{\beta}=us_{\beta}u^{-1}$ for some finite string $u=u(\alpha,\beta)$.
In particular, the two substitution rules (\ref{AB_symmetric}) and (\ref{AB_asymmetric}) both correspond to the substitution matrix
\begin{align}
		M_{\tau_\Lambda(p,q)}=\mqty(-1\;&\;q-2\\2-p\;&\;(p-2)(q-2)-1).
\end{align}

The corresponding induced map $\tau_Q(p,q):Q(S)\rightarrow Q(S')$ is also an inflation rule on two unit cell types whenever $S$ is convex. The rule $\tau_Q(p,q)$ can be extracted by combining $\tau_\Lambda(p,q)$ with the cell-to-letter mapping. When $q>3$:
\begin{subequations}
  \begin{eqnarray}
    1&\mapsto&1^{\frac{q-4}{2}}(2\;1^{q-3})^{p-3}\,2\;1^{\frac{q-4}{2}}, \\ 
    2&\mapsto&1^{\frac{q-4}{2}}(2\;1^{q-3})^{p-4}\,2\;1^{\frac{q-4}{2}},  \end{eqnarray}
\end{subequations}
and when $q=3$:
\begin{subequations}
  \begin{eqnarray}
	2&\mapsto&3^{\frac{1}{2}}2^{p-5}3^{\frac{1}{2}}, \\
	3&\mapsto&3^{\frac{1}{2}}2^{p-6}3^{\frac{1}{2}}. 
  \end{eqnarray}
\end{subequations}
The matrices $M_{\tau_Q(p,q)}$ and $M_{\tau_\Lambda(p,q)}$ are related by change of basis. They have unit determinant, and eigenvalues 
\begin{align}
\label{eq:lambda}
\lambda_{\pm}(p,q)=\gamma_{p,q}\pm(\gamma_{p,q}^{2}-1)^{1/2}
\end{align}
with 
\begin{align}
\label{eq:gamma}
\gamma_{p,q}:=\frac{(p-2)(q-2)}{2}-1.
\end{align}
These results are consistent with the previous findings of Ref.~\cite{Rietman}. Note that the largest eigenvalue $\lambda_{+}(p,q)$ is irrational and PV whenever $\frac{1}{p}+\frac{1}{q}<\frac{1}{2}$. When $S$ is nonconvex, $\tau_Q(p,q)$ acts on the many cell types of $Q(S)$ in a complicated way. However, every tile set $S$ maps to a convex set under finitely many vertex completions, and vertex completion preserves convexity once established.

\section{Conformal quasicrystals on the boundary of the disk}
 
Consider any finite simply-connected tile set $S$ of the $\qty{p,q}$ tessellation. By repeatedly carrying out layers of vertex completions we generate a sequence of tile sets $\qty{S_n}_{n\in\mathbb{Z}_{\ge 0}}$, with initial element $S_0=S$, and $\qty{Q(S_n)}_{n\in\mathbb{Z}_{\ge 0}}$. Concurrently we can consider a family of holographic TN models defined on each $S_n$ whose DOF live on the boundary of $S_n$. In the limit $n\rightarrow\infty$ the tile set covers the entire hyperbolic disk and we can ask about the unit cell assignment $Q_\infty(S):=\lim_{n\rightarrow\infty}Q(S_n)$ living at the disk boundary. We can interpret $Q_\infty(S)$ as a type of emergent quasicrystal harboring the DOF. To see this, recall that for all $n$ above some $n^*>0$, $Q(S_n)$ contains only two cell types and thus $\tau_Q(p,q)$ acts on these two cell types as an inflation rule. Taking $Q(S_{n^*})$ as an initial string and iterating $\tau_Q(p,q)$ infinitely many times generates the infinite cyclically-ordered string $Q_\infty(S)$ which is self-same under $\tau_Q(p,q)$. Thus $Q_\infty(S)$ is locally isomorphic to a $\tau_Q(p,q)$-quasicrystal, which is the promised quasicrystalline interpretation. We call any $Q_\infty(S)$ obtained in this manner a $\qty{p,q}$ \textit{conformal quasicrystal} (CQC).

\section{Discrete conformal geometry}
 
Our construction naturally endows CQCs with a discrete analog of conformal geometry. In the continuum, conformal geometry refers to those properties of a (pseudo-)Riemannian manifold that are invariant under position-dependent rescalings (or \textit{Weyl transformations}) of the metric tensor $g_{\mu\nu}(x)\mapsto \Omega^{2}(x)g_{\mu\nu}(x)$. Thus, length scales are not well-defined in conformal geometry. Conformal properties are preserved under \textit{conformal maps}, diffeomorphisms fixing the metric up to a Weyl transformation. These are all the diffeomorphisms in $d=1$ and just the angle-preserving diffeomorphisms in $d\ge 2$. In physics, conformal geometries underlie \textit{conformal field theories} (CFTs), which are quantum field theories that are symmetric under such conformal maps \cite{Ginsparg:1988ui, DiFrancesco:1997nk}. Here, we show that one can relate any two CQCs with the same $\qty{p,q}$ via a discrete analog of a Weyl transformation. Hence CQCs inherit discrete conformal geometry, defined to be the properties fixed under discrete Weyl transformations.

Let $M(S)\not=S$ denote a finite simply-connected tile set obtained from $S$ in any manner. $Q_\infty(S)$ and $Q_\infty(M(S))$ are locally isomorphic, but they differ in their global structures. To relate $Q_\infty(S)$ and $Q_\infty(M(S))$, a map $\Omega$ such that $\Omega(Q_\infty(S))=Q_\infty(M(S))$ is needed. We say that any such $\Omega$ is a \textit{discrete Weyl transformation}. $\Omega$ acts upon the global structure of the CQC by a position-dependent set of inflations and deflations. The TN model is correspondingly transformed by adjoining or removing tensors from the edges in a position dependent way.

The intuition behind our definition comes from an analogy with Weyl transformations in continuum AdS/CFT. There, a Weyl transformation corresponds to a change of choice of how to approach the boundary \cite{KaplanNotes}. More precisely, the line element for $(d+1)$-dimensional hyperbolic space can be written in global co-ordinates as
	\begin{align}
		\textrm{d}s^{2}=\frac{\textrm{d}\rho^2+\sin^2(\rho)\textrm{d}\Omega_{d}^{2}}{\cos^2(\rho)},
	\end{align}
	where $\rho\in [0,\pi/2)$ is the radial co-ordinate and $\textrm{d}\Omega_{d}^{2}$ is the line element on the unit $d$-sphere $S_{d}$.  Now consider a family of $d$-dimensional hypersurfaces of spherical topology, parameterized by $\rho(x,\epsilon)=\frac{\pi}{2}-\epsilon f(x)$ where $x$ are the co-ordinates on $S_{d}$ and $f(x)$ is an arbitrary positive function on $S_{d}$. To approach the boundary, we take $\epsilon\rightarrow 0^+$ in an $x$-independent way.  As $\epsilon\rightarrow 0^+$, a constant-$\epsilon$ hypersurface has induced line element:
	\begin{align}
		\textrm{d}s_{\epsilon}^{2}(x)\sim\frac{2}{\epsilon^{2}}\frac{\textrm{d}\Omega_d^{2}(x)}{f^2(x)},
	\end{align}
	so we see that changing the hypersurface profile $f(x)\mapsto \Omega^{-1}(x)f(x)$ induces the Weyl transformation 
	\begin{align}
		\textrm{d}s_{\epsilon}^{2}(x)\mapsto \Omega^{2}(x) \textrm{d}s_{\epsilon}^{2}(x).
	\end{align}
	In our construction, the choice of initial tile set $S$ is analogous to the choice of hypersurface profile $f(x)$, while iterating vertex completions infinitely many times corresponds to taking $\epsilon\rightarrow0^+$. Thus, it is natural to interpret $S\mapsto M(S)$ as inducing a Weyl-like transformation $Q_\infty(S)\mapsto \Omega(Q_\infty(S))=Q_\infty(M(S))$. 
	
\section{Symmetries of conformal quasicrystals in $d=1$}  
Now let us emphasize two different types of symmetry possessed by any CQC $Q_{\infty}(S)$ of type $\{p,q\}$.  First, it has a kind of exact scale symmetry: invariance under $\tau_{Q}(p,q)$.  Second, it is invariant, up to a discrete Weyl transformation, under an infinite discrete group called the triangle group $\Delta(2,p,q)$.  This is the symmetry group of the $\{p,q\}$ tiling; it is generated by reflections across the fundamental right-triangle domains of the tiling \cite{Magnus}.  Each such reflection induces a conformal map of the boundary into itself.  Under an element of $\Delta(2,p,q)$, $S$ maps to a different simply-connected $S'$ which produces $Q_\infty(S')$. Since $S'=M(S)$ for some $M$, it follows that $Q_\infty(S)$ and $Q_\infty(S')$ are related by a discrete Weyl transformation, as required. Thus we regard $\Delta(2,p,q)$ as a discretization (i) of the isometry group of the hyperbolic disk which acts on the bulk tessellation; and (ii) of the conformal group of the circular boundary which acts on the CQCs.

\section{Discussion}
\label{sec:Discussion}
 
Before discussing our results, let us briefly summarize: 

The past decade has seen exciting developments in our understanding of quantum gravity: in particular, arguments based on holography and gauge/gravity duality have led to a tantalizing but still-fragmentary picture in which spacetime emerges from the pattern of entanglement in an underlying quantum system.   In particular, there has been much recent interest in attempts to make this picture more concrete, 
in the context of AdS/CFT, 
by developing a discrete formulation of holography, based on replacing the continuous hyperbolic ``bulk" space by a discrete regular tesselation of hyperbolic space (or some other tesselation which respects a large discrete subgroup of the original space's symmetries).  These developments are part of a family of ideas which are sometimes summarized by the slogan ``it from qubit."  

Previous works have focused on discretizing the bulk geometry, but have not thought through the corresponding implications for the boundary geometry.  In this paper, we have shown that when one discretizes the bulk geometry in a natural way (e.g. on an regular tesselation), one {\it also} induces a remarkable discretization of the lower-dimensional ``boundary" geometry into a fascinating new kind of discrete geometric structure which we call a ``conformal quasicrystal."  The main goal of this paper was to point out the existence of these new structures, to define and explain their basic properties, and to emphasize that they appear to be the natural discrete spaces living on the boundary side of the holographic duality.  

We think this is an important clue for the ongoing efforts to formulate a discrete version of holography (perhaps in terms of tensor networks) 
\footnote{Note that, besides AdS/CFT, examples of holography include the BFSS matrix model \cite{Banks:1996vh} and the Klebanov-Strassler construction \cite{Klebanov:2000hb}.   The ideas described in this paper relate to the AdS/CFT context; but we hope that progress in this context may lead to progress in the other holographic settings as well.}:
in a correct discrete formulation of 
AdS/CFT, 
we expect the boundary theory to live on a conformal quasicrystal (rather than on an ordinary lattice, as imagined in previous works).  Similarly, since the boundary theory in 
AdS/CFT
is scale invariant, it suggests that the natural way to discretize and numerically simulate scale invariant systems (such as conformal field theories, or condensed matter systems near their critical points) is to discretize them on a conformal quasicrystal, rather than on a periodic lattice.  Thus, we hope that this work opens the door to more efficient simulation of the dynamics and quantum states of such systems (although much work remains in order to translate this hope into practice).

Let us now mention various directions for future work:

We expect our definitions and results to extend readily to higher dimensions.  Consider, for example, the self-dual $\qty{3,5,3}$ regular honeycomb in three-dimensional hyperbolic space. This is constructed by gluing together icosahedra such that the vertex figure at every corner is a dodecahedron \cite{CoxeterHyperbolic,CoxeterRegularPolytopes}.  Now imagine that we take the initial seed $S$ to be one such icosahedron, and then carry out vertex completion, assigning boundary unit cells at every step.  At each step we obtain a 2D layer of spherical topology with 12 points of exact five-fold orientational symmetry (or, more precisely, $D_{5}$ symmetry), lying radially above the 12 vertices of the initial icosahedron.  Hence, after iterating infinitely many times, we obtain a layer of spherical topology at the boundary, tesselated by an infinite number of tiles.  This layer still has 12 points of $D_{5}$ symmetry, and in the vicinity of any such point it appears to be an infinite tiling of the 2D Euclidean plane.  Since points of $D_{5}$ symmetry are forbidden in an ordinary (periodic) 2D crystal, we expect the boundary to once again be quasicrystalline. Indeed, the 2D quasicrystalline tilings with 5-fold orientational order are all closely related to the Penrose tiling~\cite{Gardner, Senechal, BaakeGrimm, Boyle:2016iey,RWM}.  (Note that, although the usual Penrose tiling has at most one point of exact $D_{5}$ symmetry, the Penrose-like CQC living at the boundary of $\qty{3,5,3}$ can have 12 such points, as explained above.  This is the higher-dimensional analogue of a phenomenon that already occurs in the 1D CQCs constructed above: the CQC generated from a regular $p$-gon initial seed will have $2p$ points of perfect reflection symmetry, in contrast to the self-similar 1D quasicrystal that it locally resembles, which has at most one such point.)

The conjecture that $\qty{3,5,3}$ hosts a CQC generalization of the Penrose tiling on its boundary is supported by a calculation \cite{Nemeth} of the ratio $N_{k+1}/N_k$ as $k\rightarrow\infty$, where $N_k$ is the number of boundary cells after $k$ layers of vertex completion (this ratio is expected to be equal to the largest eigenvalue of the matrix $M_\tau$ associated to the inflation rule $\tau$ induced by the vertex completion).  The limiting ratio is found to be $N_{k+1}/N_k\to\varphi^8$, where $\varphi$ is the golden ratio.  Note that this is irrational and PV (as expected for a 2D quasicrystal) and is a power of $\varphi$ (as expected for a quasicrystal with 5-fold orientational order in particular \cite{Gardner, Senechal, BaakeGrimm, Boyle:2016iey}).  If we compare this to the scaling ratio for the standard Penrose tiling ($N_{k+1}/N_k\to\varphi^2$), it suggests either that a single layer of vertex completion in the $\{3,5,3\}$ honeycomb corresponds to four iterations of the usual Penrose tiling inflation rule, or else that the Penrose tiling and $\{3,5,3\}$ CQC are related in some way besides local isomorphism.  


We have seen that applying vertex completion to a regular $\qty{p,q}$ tiling of hyperbolic space yields a $\tau_Q(p,q)$-quasicrystal. There are, however, other $\tau$-quasicrystals that are not obtained in this way -- like the example given above in Eq.~(\ref{tau_QC_example}).  We wonder whether every remaining inflation rule $\tau$ is also naturally induced by applying some appropriate generalization of vertex completion to some appropriate tiling of hyperbolic space. Tilings beyond the regular $\{p,q\}$ type would seem to be necessary.

The remarkable conformal geometry of CQCs suggests that their importance extends beyond the realm of holographic TN models. We propose to use them as the underlying spaces on which to discretize CFTs, in at least two ways. First, one can discretize a QFT partition function onto a $\tau$-quasicrystal directly. After implementing a real space renormalization group procedure whereby DOF are decimated according to $\tau^{-1}$, the RG fixed points can be identified with the discrete CFTs.  Second, one can work by analogy with AdS/CFT. Consider the partition function $Z_{S,n}[y]$ of a discretized field theory on the $n$-fold image $S_n$ under vertex completion of a simply-connected subset $S$ of the bulk tessellation, such that the fields take values $y$ at the boundary of $S_n$. As $n\rightarrow\infty$, we propose to interpret $Z_{S,\infty}[y]$ as the moment-generating function of a discrete CFT on the boundary.  

From a practical standpoint, such CQC-based discretizations may be important for studying scale-invariant physical systems such as CFTs or condensed matter systems at their critical points. These discretizations should permit finite-size simulations of such systems, in ways which preserve an {\it exact} discrete subgroup of their scale symmetry. This is in contrast with ordinary lattice gauge theory simulations, or other periodic lattice models, which instead preserve an exact discrete subgroup of translation symmetry, at the cost of breaking scale invariance.  We can attempt to finitize a numerical calculation on a conformal quasicrystal by restricting to an annulus in the reciprocal space of the quasicrystal and imposing the boundary condition that observables on the inner and outer boundaries of the annulus are the same up to an appropriate scale factor. This would provide the analog in discrete CFTs of a periodic boundary condition in a periodic lattice model, along the scale direction instead of in real space.  The resulting finite problem can then be attacked with a large arsenal of Monte Carlo and tensor network methods.  Our hope is that this approach may lead to more accurate and efficient numerical simulation of such systems, perhaps allowing us to simulate much larger systems, and leading to qualitative improvements in our understanding.

\begin{figure}[t]
\includegraphics[width=.48\textwidth]{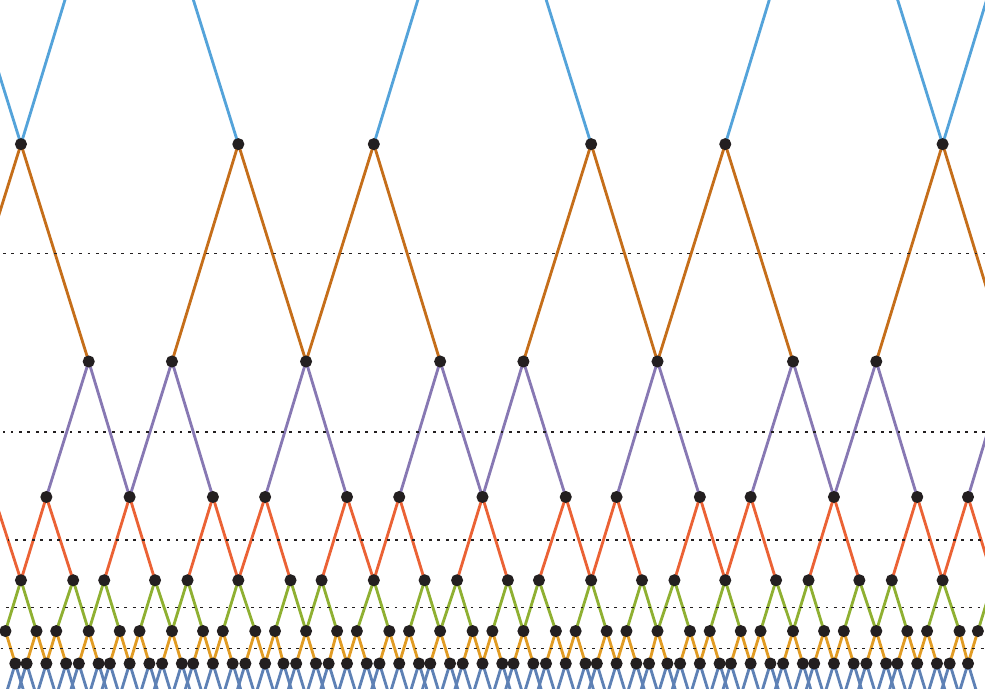}
\caption{A `quasi-MERA' tensor network based on the 1D Fibonacci quasicrystal. Vertices denote tensors and edges denote contractions between tensors. 3- and 4-legged tensors represent isometries and disentanglers, respectively. Dotted lines separate layers related by the Fibonacci deflation rule.}
\label{Fibonacci}
\end{figure} 

TN algorithms based on inflations and deflations may also find use in the numerical description of quantum critical states. We can imagine implementing a MERA-like coarse-graining scheme induced by the quasicrystalline structure. Consider the simplest $\tau$-quasicrystals, the 1D Fibonacci quasicrystals generated by the inflation rule $\tau:\{\alpha,\beta\}\mapsto\{\beta^{1/2}\beta^{1/2},\beta^{1/2}\alpha \beta^{1/2}\}$. Given an initial Fibonacci quasicrystal $\Pi$, we can use a deflation-like rule to obtain a coarser quasicrystal $c(\Pi)$ in two steps: (i) slice every $\beta$ interval into two halves, $\beta^{1/2}$, resulting in a string which can be uniquely partitioned into substrings of the form $\beta^{1/2}\beta^{1/2}$ and $\beta^{1/2}\alpha\beta^{1/2}$; (ii) glue these substrings together to form intervals $\alpha'= \beta^{1/2}\beta^{1/2}$ and $\beta'=\beta^{1/2}\alpha\beta^{1/2}$ of respective lengths $L_{\alpha'}=L_\beta$ and $L_{\beta'}=L_\alpha+L_\beta$. Iterating this procedure yields a sequence $\qty{c^k(\Pi)}_{k\in\mathbb{Z}_{\ge 0}}$ of successively coarser quasicrystals.  We can now construct a MERA-like circuit as follows. Embed each quasicrystal $c^k(\Pi)$ into the two-dimensional plane, such that $c^{k+1}(\Pi)$ is parallel to and lies above $c^k(\Pi)$ for every $k$. Assign to every point in $c^k(\Pi)$ a tensor, and contract each tensor only with its nearest neighbors in $c^{k+1}(\Pi)$ and $c^{k-1}(\Pi)$. This results in 3- and 4-legged tensors; if we choose these to be isometries and unitary disentanglers, respectively, then we acquire the desired MERA-like circuit. We call this network, depicted in Fig.~\ref{Fibonacci}, the \textit{quasi-MERA} due to its relationship with the quasicrystalline inflation rule $\tau$.  

To see why the quasi-MERA may be special, we contrast the deflation-based decimation procedure of the $\tau$-quasicrystal with the standard block decimation of the periodic crystal. Unlike the periodic crystal, for every $\tau$-quasicrystal layer there is a local refinement (inflation) rule which is inverted by an unambiguous {\it local} coarse-graining (deflation) rule. To see why this cannot hold for an ordinary crystal, consider the example of a 1D periodic lattice with intervals of length $a$. We can produce a more refined lattice (with intervals of length $a/2$) by the local rule of chopping each interval in half; but the inverse coarse-graining transformation is not locally well defined: there is an ambiguity about which pair of short intervals to glue together to make a long one, so that $N$ `joiners' at widely-separated points on the lattice would make different, incompatible choices about which tiles to join, and ${\cal O}(N)$ defects would be produced. Hence, block decimating a periodic lattice destroys information about how to locally recover the original lattice. Since the ordinary (binary) MERA architecture relies on this form of block decimation, its description of a quantum state is necessarily lossy, even at criticality. By contrast, when coarse-graining the Fibonacci quasicrystal as in Fig.~\ref{Fibonacci}, one can locally and unambiguously determine, from the tiles $\alpha$ and $\beta$ in a given layer, where to place the larger intervals $\alpha'$ and $\beta'$ in the next layer up. In this sense, the coarse-graining transformation for a $\tau$-quasicrystal is lossless, in contrast to the block decimation of a periodic lattice. For this reason, we ask whether the quasi-MERA can give an exact description of a quantum critical state provided that the system is already discretized on the appropriate quasicrystal. This is a nontrivial statement that requires numerical checking and benchmarking. 

Geometrically, the quasi-MERA construction of Fig.~\ref{Fibonacci} generates a novel emergent `tiling' of the hyperbolic half-plane by hexagons, with 3 or 4 glued together at a vertex. However, this network should be conceptually separated from the TNs considered in the main paper because, unlike those, this network can harbor unitary disentanglers and has a preferred causal direction. Specifically, we do not believe that the quasi-MERA should be interpreted as a discretization of a space-like slice of $\mathrm{AdS}_{2+1}$.  In view of Refs. \cite{Czech:2015kbp} and \cite{Milsted:2018san}, we ask whether Fig.~\ref{Fibonacci} could be interpreted as a discretization of the kinematic space, or some other causal geometry. 

We leave the investigation of these many exciting possibilities for future studies. 

\begin{acknowledgements}
We thank Ehud Altman, Guifre Vidal and Mudassir Moosa for helpful discussions.  LB acknowledges support from an NSERC Discovery Grant. MD acknowledges the NSF Graduate Research Fellowship Program and the Perimeter Institute for Theoretical Physics for partial funding. FF acknowledges support from a Lindemann Trust Fellowship of the English Speaking Union, and the Astor Junior Research Fellowship of New College, Oxford.  Research at Perimeter Institute is supported by the Government of Canada through the Department of Innovation, Science and Economic Development and by the Province of Ontario through the Ministry of Research and Innovation. Figs.~\ref{Assignment} and~\ref{inflation} have been generated with the assistance of the KaleidoTile software \cite{KaleidoTile}.
\end{acknowledgements}


\begin{thebibliography}{99}
  
\bibitem{Maldacena:1997re} 
  J.~M.~Maldacena,
  {\it The Large N limit of superconformal field theories and supergravity},
  Int.\ J.\ Theor.\ Phys.\  {\bf 38}, 1113 (1999)
  [Adv.\ Theor.\ Math.\ Phys.\  {\bf 2}, 231 (1998)]
  [hep-th/9711200].
  
\bibitem{Witten:1998qj} 
  E.~Witten,
  {\it Anti-de Sitter space and holography},
  Adv.\ Theor.\ Math.\ Phys.\  {\bf 2}, 253 (1998)
  [hep-th/9802150].
  
\bibitem{KaplanNotes}
  J.~Kaplan, {\it Lectures on Ads/CFT from the Bottom Up}, Johns Hopkins University course notes (unpublished).

\bibitem{VanRaamsdonk:2010pw} 
  M.~Van Raamsdonk,
  Gen.\ Rel.\ Grav.\  {\bf 42}, 2323 (2010)
  [Int.\ J.\ Mod.\ Phys.\ D {\bf 19}, 2429 (2010)]
  doi:10.1007/s10714-010-1034-0, 10.1142/S0218271810018529
  [arXiv:1005.3035 [hep-th]].
  
\bibitem{Wilson:1974sk} 
  K.~G.~Wilson,
  Phys.\ Rev.\ D {\bf 10}, 2445 (1974).
  doi:10.1103/PhysRevD.10.2445
  
  
\bibitem{Vidal:2007hda} 
  G.~Vidal,
  {\it Entanglement Renormalization},
  Phys.\ Rev.\ Lett.\  {\bf 99}, no. 22, 220405 (2007)
  [cond-mat/0512165].
  
\bibitem{Vidal:2008zz} 
  G.~Vidal,
  {\it Class of Quantum Many-Body States That Can Be Efficiently Simulated},
  Phys.\ Rev.\ Lett.\  {\bf 101}, 110501 (2008)
  [quant-ph/0610099].
  
\bibitem{Vidal:Intro}
  G.~Vidal,
  {\it Entanglement Renormalization: an Introduction}, in {\it Understanding Quantum Phase Transitions}, CRC Press (2010)
  [arXiv:0912.1651].
  
\bibitem{Pfeifer:2008jt} 
  R.~N.~C.~Pfeifer, G.~Evenbly and G.~Vidal,
  {\it Entanglement renormalization, scale invariance, and quantum criticality},
  Phys.\ Rev.\ A {\bf 79}, 040301 (2009)
  [arXiv:0810.0580 [cond-mat.str-el]].
  
\bibitem{EvenblyVidalTNgeometry}
  G.~Evenbly and G.~Vidal,
  {\it Tensor Network States and Geometry}, 
  J.\ Stat.\ Phys.\ {\bf 145}, 891 (2011)
  [arXiv:1106.1082 [quant-ph]].
  
\bibitem{Orus:2014poa} 
  R.~Orus,
  {\it Advances on Tensor Network Theory: Symmetries, Fermions, Entanglement, and Holography},
  Eur.\ Phys.\ J.\ B {\bf 87}, 280 (2014)
  [arXiv:1407.6552 [cond-mat.str-el]].
  
  
\bibitem{Swingle:2009bg} 
  B.~Swingle,
  {\it Entanglement Renormalization and Holography},
  Phys.\ Rev.\ D {\bf 86}, 065007 (2012)
  [arXiv:0905.1317 [cond-mat.str-el]].
  
\bibitem{Swingle:2012wq} 
  B.~Swingle,
  {\it Constructing holographic spacetimes using entanglement renormalization},
  arXiv:1209.3304 [hep-th].
  
\bibitem{Qi:2013caa} 
  X.~L.~Qi,
  {\it Exact holographic mapping and emergent space-time geometry},
  arXiv:1309.6282 [hep-th].

\bibitem{MolinaVilaplana:2012nz} 
  J.~Molina-Vilaplana,
  {\it Holographic Geometries of one-dimensional gapped quantum systems from Tensor Network States},
  JHEP {\bf 1305}, 024 (2013)
  [arXiv:1210.6759 [hep-th]].
  
\bibitem{Pastawski:2015qua} 
  F.~Pastawski, B.~Yoshida, D.~Harlow and J.~Preskill,
  {\it Holographic quantum error-correcting codes: Toy models for the bulk/boundary correspondence},
  JHEP {\bf 1506}, 149 (2015)
  [arXiv:1503.06237 [hep-th]].
  
\bibitem{Hayden:2016cfa} 
  P.~Hayden, S.~Nezami, X.~L.~Qi, N.~Thomas, M.~Walter and Z.~Yang,
  {\it Holographic duality from random tensor networks},
  JHEP {\bf 1611}, 009 (2016)
  [arXiv:1601.01694 [hep-th]].

\bibitem{Czech:2015kbp} 
  B.~Czech, L.~Lamprou, S.~McCandlish and J.~Sully,
  {\it Tensor Networks from Kinematic Space},
  JHEP {\bf 1607}, 100 (2016)
  [arXiv:1512.01548 [hep-th]].
 
\bibitem{Han:2016xmb} 
  M.~Han and L.~Y.~Hung,
  {\it Loop Quantum Gravity, Exact Holographic Mapping, and Holographic Entanglement Entropy},
  Phys.\ Rev.\ D {\bf 95}, no. 2, 024011 (2017)
  [arXiv:1610.02134 [hep-th]].
  
\bibitem{Qi:2017ohu} 
  X.~L.~Qi, Z.~Yang and Y.~Z.~You,
  {\it Holographic coherent states from random tensor networks},
  JHEP {\bf 1708}, 060 (2017)
  [arXiv:1703.06533 [hep-th]].
  
\bibitem{Evenbly:2017hyg} 
  G.~Evenbly,
  {\it Hyperinvariant Tensor Networks and Holography},
  Phys.\ Rev.\ Lett.\  {\bf 119}, no. 14, 141602 (2017)
  [arXiv:1704.04229 [quant-ph]].
  
\bibitem{Osborne:2017woa} 
  T.~J.~Osborne and D.~E.~Stiegemann,
  {\it Dynamics for holographic codes},
  arXiv:1706.08823 [quant-ph].
  
\bibitem{Bao:2017qmt} 
  N.~Bao, C.~Cao, S.~M.~Carroll and A.~Chatwin-Davies,
  {\it De Sitter Space as a Tensor Network: Cosmic No-Hair, Complementarity, and Complexity},
  Phys.\ Rev.\ D {\bf 96}, no. 12, 123536 (2017)
  [arXiv:1709.03513 [hep-th]].
  
\bibitem{Jahn:2017tls} 
  A.~Jahn, M.~Gluza, F.~Pastawski and J.~Eisert,
  {\it Holography and criticality in matchgate tensor networks},
  arXiv:1711.03109 [quant-ph].
  

\bibitem{NielsenChuang}
    M.A.~Nielsen and I.L.~Chuang, {\it Quantum computation and quantum information}, Cambridge University Press (2010).
  
  
\bibitem{Czech:2015qta} 
  B.~Czech, L.~Lamprou, S.~McCandlish and J.~Sully,
  {\it Integral Geometry and Holography},
  JHEP {\bf 1510}, 175 (2015)
  [arXiv:1505.05515 [hep-th]].
 
\bibitem{Milsted:2018san} 
  A.~Milsted and G.~Vidal,
  arXiv:1812.00529 [hep-th].
 
\bibitem{CoxeterHyperbolic}
  H.S.M.~Coxeter, 
  {\it Regular honeycombs in hyperbolic space},
  Proceedings of the International Congress of Mathematicians {\bf 3}, 155 (1954).  
  
\bibitem{Brekke:1993gf} 
  L.~Brekke and P.~G.~O.~Freund,
  Phys.\ Rept.\  {\bf 233}, 1 (1993).
  doi:10.1016/0370-1573(93)90043-D
  
\bibitem{luck1993classification}
  J.M.~Luck, {\it A classification of critical phenomena on quasicrystals and other aperiodic structures}, Europhysics Letters 
  {\bf 24}, 359 (1993).
  
\bibitem{Senechal}
  M.~Senechal, {\it Quasicrystals and Geometry}, Cambridge University Press (1995).
  
\bibitem{Janot}
  C.~Janot, {\it Quasicrystals: A Primer}, Clarendon Press, Oxford (1994).
  
\bibitem{BaakeGrimm}
  M.~Baake and U.~Grimm, 
  {\it Aperiodic Order: Volume 1: A Mathematical Invitation},
  Cambridge University Press (2013).
  
\bibitem{Boyle:2016sjm} 
  L.~Boyle and P.~J.~Steinhardt,
  {\it Self-Similar One-Dimensional Quasilattices},
  arXiv:1608.08220 [math-ph].

\bibitem{Boyle:2016iey} 
  L.~Boyle and P.~J.~Steinhardt,
  {\it Coxeter Pairs, Ammann Patterns and Penrose-like Tilings},
  arXiv:1608.08215 [math-ph].
  
\bibitem{Gardner}  
  M.~Gardner, 
  {\it Extraordinary nonperiodic tiling that enriches the theory of tiles},
  Sci.\ Amer.\ {\bf 236}, 110 (1977).
  
\bibitem{BombieriTaylor}
  E.~Bombieri and J.E.~Taylor, 
  {\it Which distributions of matter diffract?  An initial investigation}, J.\ Phys.\ Colloques {\bf 47}, C3-19 (1986);
  {\it Quasicrystals, tilings, and algebraic number theory: some preliminary connections}, Contemp.\ Math {\bf 64},
  241 (1987).
  
\bibitem{CoxeterRegularPolytopes}
  H.S.M.~Coxeter, {\it Regular Polytopes}, Methuen \& Co. Ltd., London (1948).
  
\bibitem{Rietman}
  R.~Rietman, B.~Nienhuis, and J.~Oitmaa,
  {\it The Ising model on hyperlattices},
  Journal of Physics A: Mathematical and General {\bf 25}, 6577 (1992).
  doi:10.1088/0305-4470/25/24/012  
  
\bibitem{Ginsparg:1988ui} 
  P.~H.~Ginsparg,
  {\it Applied Conformal Field Theory}, in {\it Fields, Strings and Critical Phenomena} (Les Houches, Session XLIX, 1988) 
  [hep-th/9108028].

\bibitem{DiFrancesco:1997nk} 
  P.~Di Francesco, P.~Mathieu and D.~Senechal,
  {\it Conformal Field Theory}, Springer-Verlag, New York (1997).

\bibitem{Magnus}
  W.~Magnus, 
  {\it Noneuclidean Tesselations and Their Groups}, 
  Academic Press, New York and London (1974).

\bibitem{Banks:1996vh} 
  T.~Banks, W.~Fischler, S.~H.~Shenker and L.~Susskind,
  Phys.\ Rev.\ D {\bf 55}, 5112 (1997)
  doi:10.1103/PhysRevD.55.5112
  [hep-th/9610043].
  
\bibitem{Klebanov:2000hb} 
  I.~R.~Klebanov and M.~J.~Strassler,
  JHEP {\bf 0008}, 052 (2000)
  doi:10.1088/1126-6708/2000/08/052
  [hep-th/0007191].

\bibitem{RWM}
  D.S.~Rokhsar, D.C.~Wright,and N.D.~Mermin, {\it The two-dimensional quasicrystallographic space groups with rotational symmetries less than 23-fold}, Acta Cryst.\ A {\bf 44}, 197 (1988).

\bibitem{Nemeth}
  L.~Nemeth, 
  {\it The growing ratios of hyperbolic regular mosaics with bounded cells}, 
  Armenian J. of Math. {\bf 9}, 1 (2017) 
  [arXiv:1705.02648 [math.MG]].

\bibitem{KaleidoTile}
  J.~Weeks, 
  {\it KaleidoTile}, 
  a computer program for creating spherical, Euclidean and hyperbolic tilings.
  

\end{thebibliography}
\end{document}